\documentclass[twocolumn,showpacs,preprintnumbers,amsmath,amssymb,prb,aps]{revtex4-1}
\usepackage{epsfig}
\usepackage{epstopdf}
\usepackage{graphicx}
\usepackage{dcolumn}
\usepackage{bm}
\usepackage{textcomp}
\usepackage{array}
\usepackage{subcaption}
\usepackage{booktabs}

\begin{document}

\title{Achieving high \textit{figure-of-merit} in Nb-doped Na$_{0.74}$CoO$_{2}$ compound at high temperature region}
\author{Arzena Khatun$^{1,}$}
\altaffiliation{Electronic mail: khatunarzena@gmail.com}
\author{Shamim Sk$^{1}$}
\author{Jayashree Pati$^{2}$}
\author{R. S. Dhaka$^{2}$}
\author{Sudhir K. Pandey$^{3,}$}
\altaffiliation{Electronic mail: sudhir@iitmandi.ac.in}
\affiliation{$^{1}$School of Basic Sciences, Indian Institute of Technology Mandi, Kamand - 175005, India}
\affiliation{$^{2}$Department of Physics, Indian Institute of Technology Delhi, Hauz Khas, New Delhi - 110016, India}
\affiliation{$^{3}$School of Engineering, Indian Institute of Technology Mandi, Kamand - 175005, India}

\date{\today}

\begin{abstract}
We report the thermoelectric (TE) properties of Na$_{0.74}$Co$_{0.95}$Nb$_{0.05}$O$_{2}$ in the temperature
range $300-1200$ K, as a potential candidate for p-type thermoelectric material. The experimental values of Seebeck coefficient (S) are $ \sim $  $82-121$ $ \mu $V/K measured in the temperature range $300-620$ K. The positive values of S in the entire temperature range indicates p-type behaviour of the compound. At 300 K the experimental value of thermal conductivity ($ \kappa  $) is $ \sim $ 1.88 W/m-K that increases up to $ \sim $ 420 K, then decreases till 620 K with corresponding value $ \sim $ 1.86 W/m-K. To understand the experimentally observed transport properties, we have calculated S and $\rho$ of this compound. Then, based on theoretical understanding, we have estimated \textit{figure-of-merit} (ZT) up to 1200 K by using calculated S and $\rho$ values with extrapolated experimental $\kappa$. The value of ZT is found to be $\sim$ 0.03 at 300 K, whereas, the highest value is observed as $\sim$ 1.7 at 1200 K. Finally, we have calculated the efficiency ($\eta$) by keeping the cold end temperature (T$_{c}$) fixed at 500 K and varying hot end temperature (T$_{h}$) from 500 to 1200 K, respectively. The maximum value of $\eta$ is found to be $\sim$ 8 $ \% $, when T$_{c}$ and T$_{h}$ are fixed at 500 and 1200 K, respectively. This result suggests that Na$_{0.74}$Co$_{0.95}$Nb$_{0.05}$O$_{2}$ compound can be used as a p-leg for making high temperature TE generator (TEG).  


\end{abstract}

\maketitle

Recently clean energy technologies become the world demand because they are pollution free and also participate in energy problem solution. Today's world runs on a series of electrical reactions. Various sources of energy such as solar, wind, hydro-power, biomass and thermal energy are used to generate electricity. But due to the poor conversion efficiency of engines, more than 60$ \% $  of the energy expenditure from different energy sources is directly released to the atmosphere in the form of waste heat.\cite{PE,nF} The amount of waste heat is increasing day by day that changes global climate and becoming alarming for environment. Therefore, to fulfill the requirement of electricity without affecting environmental conditions necessitate the discovery of new clean energy sources. The thermoelectric (TE) materials are one of the most promising candidates in clean energy technology and are also expected to play a vital role in future energy systems. TE materials have the capability to convert waste heat into useful electrical energy\cite{WE2} through thermoelectric power. But all TE materials are not suitable for commercial applications. The efficiency can express the applicability of any device for commercial applications. For TE materials the efficiency is measured by a dimensionless parameter called the \textit{figure-of-merit }(ZT)\cite{zt1} 
\begin{equation}
ZT = \dfrac{S^{2}T}{\rho\kappa},
\end{equation}
where, S is Seebeck coefficient, $ \rho $ is electrical resistivity, T is absolute temperature and $ \kappa  $($=\kappa_{e} + \kappa_{ph}$) is thermal conductivity in which electronic ($\kappa_{e}$) and phononic ($\kappa_{ph}$) parts are involved. For good TE materials the magnitude of ZT should be large. The  above expression suggests that to get large magnitude of ZT the value of S should be high and the value of $ \rho $ as well as $ \kappa $ should be low. But the optimization of S, $ \rho $ and $ \kappa $ to get a high value of ZT is a challenging task because they are interrelated with each other.\cite{in1,in2} However, researchers are working hard to overcome this difficulty for improving the magnitude of ZT.    

Till now, the excellent examples of commercially available TE materials are bismuth (and lead) telluride and their alloys having highest ZT of about 1 near room temperature.\cite{Be1,Be2} But, difficulty arises in high temperature region because they are easily decomposed and oxidised at higher temperature that makes them inapplicable for high temperature applications. Oxides are good candidates for thermoelectric applications due to their comparatively lower cost, non-toxic, simple synthesis procedure and the anti-oxidization in oxidized atmosphere for long time even at elevated temperature.\cite{lt2,lt3} Relatively, metal oxides such as ZnO\cite{zn2}, Ca-Co-O\cite{ca2} and Na$ _{x} $CoO$ _{2} $\cite{na2} systems have been revealed as an incentive TE materials. 

The scientific community is showing considerable interest on Na$ _{x} $CoO$ _{2} $ because of its peculiar properties. Na$ _{x} $CoO$ _{2} $ has attracted much attention due to its remarkable TE power, high ZT\cite{rp} and the discovery of superconductivity at 5 K in the hydrate sodium cobaltate.\cite{sc} Further, Na content has a crucial role in TE properties of Na$ _{x} $CoO$ _{2} $. The larger amount of Na doping (x$ \sim $0.7) in Na$ _{x} $CoO$ _{2} $ system gives high thermopower above 75 $ \mu $V/K.\cite{19,20,21,22} In the past two decades, many researchers have studied the TE properties of Na$ _{x} $CoO$ _{2} $ in low temperature region (0-300 K).\cite{19,23,21,25,26} A combined experimental and theoretical study on TE properties of Na$ _{0.74} $CoO$ _{2} $ compound has been reported by Sk \textit{et al.}\cite{27} in $300-1200$ K. They have calculated S values at chemical potential, $\mu$ = 220 meV. Their result suggests the possibility of enhancing the value of S along with electrical conductivity ($\sigma$) by doping. Moreover, it has been often observed that doping also decreases the value of $ \kappa $.\cite{28} Thus, it is expected that through appropriate doping the value of ZT can be increased for Na$ _{0.74} $CoO$ _{2} $ compound. To look for this possibility we have doped 5 $\%$ of the Niobium (Nb) at Co site.    
 
In this  work, we have experimentally studied the TE properties (S, $ \kappa $) of Na$ _{0.74} $Co$ _{0.95} $Nb$ _{0.05} $O$ _{2} $ compound in the temperature range 300 to 620 K. At 300 K the measured value of S is $ \sim $ 82 $ \mu $V/K, whereas, the highest value of S is obtained as $ \sim $ 121 $ \mu $V/K at 620 K. The observed value of $ \kappa $ is $ \sim $ 1.88 W/m-K at 300 K then increases up to $ \sim $ 420 K. After that the values of $ \kappa $ decrease slowly till 620 K and the lowest value is found to be  $ \sim $ 1.86 W/m-K at 620 K. In order to understand the experimentally observed transport properties, we have calculated S and $\rho$ of this compound and compared with experiment. Then, we have calculated ZT using experimental S and $\kappa$ values with calculated $\rho$ value. At room temperature the observed value of ZT is found to be $\sim$ 0.03, which reaches $\sim$ 0.16 at 620 K. Based on the computational understanding, we have estimated the ZT values up to 1200 K by using calculated S and $\rho$ values with extrapolated experimental $\kappa$ values. The highest value of ZT is found to be $\sim$ 1.7 at 1200 K. Finally, we have calculated the temperature dependent efficiency ($\eta$) of Na$_{0.74}$Co$_{0.95}$Nb$_{0.05}$O$_{2}$ compound. The maximum $\eta$ is observed as $\sim$ 8 $ \% $, when cold end temperature (T$_{c}$) and hot end temperature (T$_{h}$) are fixed at 500 and 1200 K, respectively. The observed value of $\eta$ indicates that Na$_{0.74}$Co$_{0.95}$Nb$_{0.05}$O$_{2}$ can be used as a p-type material for making high temperature TEG. 

Na$ _{0.74} $Co$ _{0.95} $Nb$ _{0.05} $O$ _{2} $ compound was prepared by the conventional solid state reaction. The more details of synthesis procedure and characterization can be found in Ref.[25]. The measurements of temperature dependent S and $ \kappa $ were carried out by using home-made experimental setup.\cite{30,31,32} The measurements were conducted in the temperature range 300 to 620 K. The sample with pellet form of 8 ($ \pm $0.1) mm diameter and 1 ($ \pm $0.02) mm thickness was used for performing the measurements. 

\begin{figure}
\includegraphics[width=0.76\linewidth, height=11.0cm]{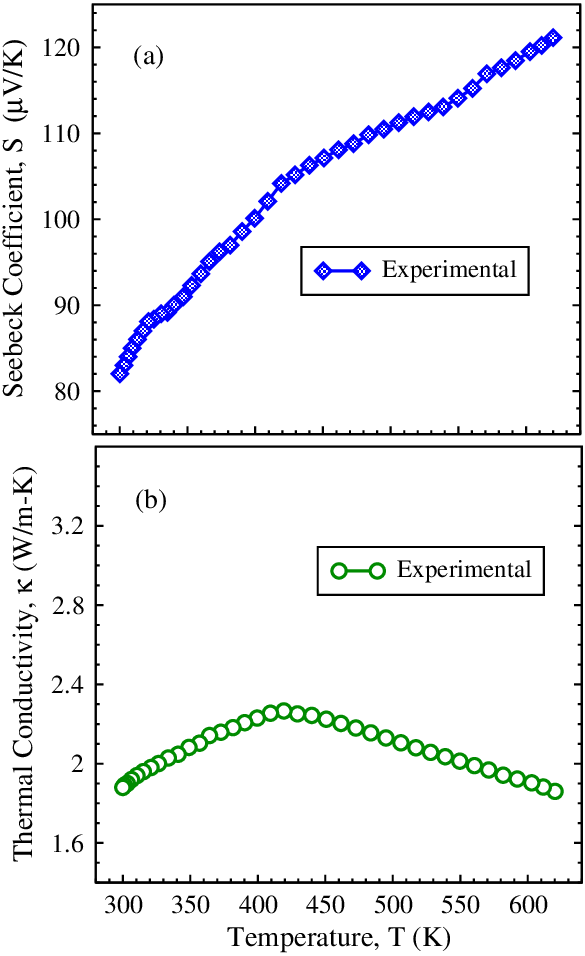} 
\caption{\small{Temperature dependence of (a) Seebeck coefficient (S) and (b) Thermal conductivity ($ \kappa $) of Na$_{0.74}$Co$ _{0.95} $Nb$ _{0.05} $O$ _{2} $.}}
\end{figure}

\begin{figure} 
\includegraphics[width=0.76\linewidth, height=5.5cm]{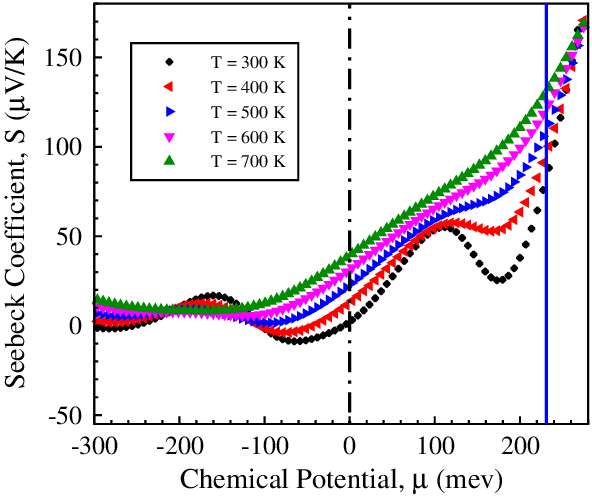} 
\caption{\small{Variation of Seebeck coefficient (S) with chemical potential ($\mu $) for Na$_{0.74}$CoO$ _{2} $ compound at various temperature.}}
\end{figure}

Fig. 1(a) presents the results of experimentally measured temperature dependent S of Na$ _{0.74} $Co$ _{0.95} $Nb$ _{0.05} $O$ _{2} $ compound. The measurement of S was carried out in the temperature range 300 to 620 K. The observed S values are positive throughout the temperature range that indicates p-type conductive transport properties of the compound. At room temperature the measured value of S is $ \sim $ 82 $ \mu $V/K, which is greater than the reported S of $ \sim $ 64 $ \mu $V/K for parent compound.\cite{27} The S vs T plot shows that S increases monotonically with increasing temperature. The highest value of S is obtained as $ \sim $ 121 $ \mu $V/K at 620 K and found to be larger than the reported value of S ($ \sim $ 118 $ \mu $V/K) for parent compound. The overall values of S in the entire temperature range are greater than the parent, Na$ _{0.74} $CoO$ _{2} $ compound. ZT of the TE material is dependent on square of S values, therefore observation of increment in the magnitude of S with increasing temperature suggests that this compound can be used for high temperature TE applications.

\begin{figure*} 
\includegraphics[width=0.8\linewidth, height=12.0cm]{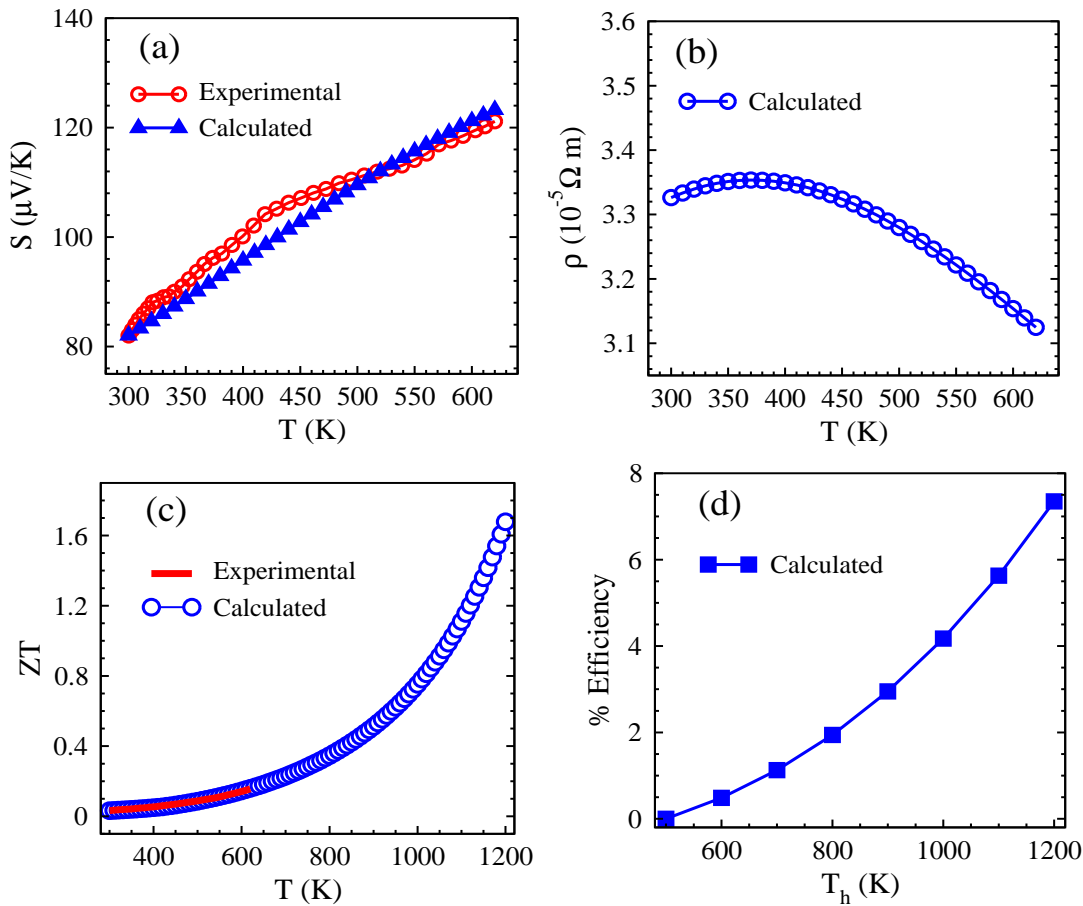} 
\caption{\small{(a) Comparison of experimental and calculated values of Seebeck coefficient (S), (b) Calculated values of resistivity ($\rho$) as a function of temperature, (c) Temperature dependence of \textit{figure-of-merit} (ZT) and (d) Efficiency ($\eta$) of Na$_{0.74}$Co$ _{0.95} $Nb$ _{0.05} $O$ _{2} $ with hot end temperature, keeping cold end temperature fixed at 500 K.}}
\end{figure*}

Fig. 1(b) shows the $ \kappa $ of Na$ _{0.74} $Co$ _{0.95} $Nb$ _{0.05} $O$ _{2} $ compound as a function of temperature. The measurement was conducted from room temperature  to  620 K. At 300 K the observed value of $ \kappa $ is $ \sim $ 1.88 W/m-K. This value of $ \kappa $ is small in comparison to the $ \kappa $ value ($ \sim $ 2.2 W/m-K) reported for parent compound.\cite{27} With increasing temperature the $ \kappa $ value increases up to $ \sim $ 420 K then decreases till the highest temperature studied. At 620 K the observed value of $ \kappa $ is found to be  $ \sim $ 1.86 W/m-K, which is smaller than the reported value of $ \kappa $ ($ \sim $ 2.4 W/m-K) measured for parent compound. In comparison with parent compound the  $ \kappa $ values  are  lower in the entire temperature range. The lower values of $ \kappa $ under the temperature range is a good signature for efficient TE materials. The lower value of $ \kappa $  helps in getting higher value of ZT and for good TE materials that is one of the most crucial requirement. 

In order to understand the experimentally observed transport properties, we have calculated S and $\rho$ with the help of calculations for parent compound Na$_{0.74}$CoO$_{2}$ reported earlier. \cite{27} In that work S was calculated by varying $\mu$ at different temperatures. Here, we have shown S vs $\mu$ plot in Fig. 2, where, $\mu$ is taken in the range -300 to 280 meV and the temperatures are varying from 300 to 700 K. We found that at $\mu$ = 231 meV, the calculated value of S is $\sim$ 82 $\mu$V/K at 300 K, which gives good match with our experimentally observed value of $\sim$ 82 $\mu$V/K for Na$ _{0.74} $Co$ _{0.95} $Nb$ _{0.05} $O$ _{2} $. At $\mu$ = 231 meV, we have calculated S using two current model\cite{33,35,36} in the temperature range $300-620$ K as shown in Fig. 3(a). From the figure it is clearly seen that calculated values of S are giving good match with experiment in the entire temperature region. The same value of $ \mu $ is used for rest of the calculations in this work.

Temperature dependent $\rho$ is also calculated using two current model\cite{Rho} for Na$_{0.74}$Co$_{0.95}$Nb$_{0.05}$O$_{2}$ compound. Fig. 3(b) presents the calculated $\rho$ as a function of temperature. To calculate the value of $ \rho $, we have used the same value of relaxation time, $\tau$ = 0.7$\times$10$^{-14}$ s as used for parent compound.\cite{27} At room temperature the value of $ \rho $ is $\sim$ 3.33$\times$10$^{-5}$ $ \Omega $ m. As the temperature increases the value of $\rho$ increases up to 370 K then decreases till the highest studied temperature. The value of $\rho$ is obtained as $\sim$ 3.12$\times$10$^{-5}$ $\Omega$ m at 620 K.  

To see the applicability of Na$_{0.74}$Co$_{0.95}$Nb$_{0.05}$O$_{2}$ compound for TE applications, we have calculated the ZT values in temperature region $300-620$ K as shown in Fig. 3(c). To calculate the ZT values, experimental S and $\kappa$ values with calculated $\rho$ value were used. The ZT vs T plot shows that ZT values increase with increasing temperature. At 300 K the observed value of ZT is found to be $\sim$ 0.03 that is larger than the ZT value ($\sim$ 0.02) of parent compound.\cite{27} With increasing temperature the ZT value reaches $\sim$ 0.16 at 620 K. At 620 K the observed value is greater than the reported value ($\sim$ 0.13) of parent compound. The overall values of ZT in the entire temperature range is greater than the parent compound. This increasing nature of ZT with temperature suggests to see the ZT value at higher temperature region where performing the experiment is not accessible in the present study.

On the basis of computational understanding, we have estimated the ZT values up to 1200 K for Na$_{0.74}$Co$_{0.95}$Nb$_{0.05}$O$_{2}$ as displayed in Fig. 3(c). Here, ZT is calculated by taking calculated values of S and $\rho$ with extrapolated experimental $\kappa$ values. The value of ZT at 300 is found to be $\sim$ 0.03, whereas, it reaches $\sim$ 1.7 at 1200 K. From the figure it is clear that ZT values are increasing with increasing temperature. The values of ZT at higher temperature region are larger than the ZT value of commercially used materials at low temperature.\cite{Be1,Be2} It is important to note that the values of ZT at higher temperature region are calculated by using extrapolated experimental $\kappa$ values and constant  $\tau$ value which is caculated at 300 K. But, $\tau$ is temperature dependent\cite{37} quantity. Therefore, by considering the proper values of temperature dependent $\tau$ and $\kappa$ at high temperature region, we believe that one may still get ZT value $\geq$ 1 at 1200 K. This result suggests that Na$_{0.74}$Co$_{0.95}$Nb$_{0.05}$O$_{2}$ compound can be used for high temperature TE applications.    

Finally, we have calculated $\eta$ of Na$_{0.74}$Co$_{0.95}$Nb$_{0.05}$O$_{2}$ compound as shown in Fig. 3(d). As we know, for making TEG both p and n-type materials are required. Our compound is p-type that is indicated by the positive values of S throughout the temperature range. In order to check the efficacy of this compound, we have calculated $\eta$ at higher temperature region by using the segmentation method.\cite{27,38} If we observe Fig. 3(c), ZT values are less in low temperature region. Therefore, we have chosen the cold end temperature as 500 K to calculate $\eta$ by varying hot end temperature from 500 to 1200 K. The maximum $\eta$ is observed as $\sim$ 8 $ \% $ when T$_{c}$ and T$_{h}$ are fixed at 500 and 1200 K, respectively. The calculated value of $\eta$ suggests that this compound can be used as a p-type material for making high temperature TEG.   

In conclusion, we have experimentally measured temperature dependent TE properties (S and $\kappa$) of Na$_{0.74}$Co$ _{0.95} $Nb$ _{0.05} $O$ _{2} $ compound in the temperature range 300 to 620 K. The observed value of S at 300 K  is $ \sim $ 82 $ \mu $V/K and at 620 K the value is found to be $ \sim $ 121 $ \mu $V/K . The values of S in the entire temperature range are positive that indicates p-type conductive transport properties of the compound. At room temperature the  observed value of $ \kappa $ is $ \sim $ 1.88 W/m-K that increases up to $ \sim $ 420 K then decreases till highest temperature studied. The lowest value of $ \kappa $ is found to be  $ \sim $ 1.86 W/m-K at 620 K. The low values of $ \kappa $ under the temperature range help in getting higher value of ZT. To understand the experimentally observed transport properties, we have calculated S and $\rho$ of this compound. By taking experimental S and $\kappa$ values with calculated $\rho$ value we have calculated ZT. The highest value of ZT is found to be $\sim$ 0.16 at 620 K. Then, we have estimated the ZT values up to 1200 K on the basis of theoretical understanding. For this we have used calculated S and $\rho$ values with extrapolated experimental $\kappa$ values. At 300 K the value of ZT is $\sim$ 0.03, which reaches $\sim$ 1.7 at 1200 K. Finally, we have calculated $\eta$ of Na$_{0.74}$Co$_{0.95}$Nb$_{0.05}$O$_{2}$ by fixing T$_{c}$ at 500 K and varying T$_{h}$ from 500 to 1200 K. The maximum value of  $\eta$ is observed as $\sim$ 8 $ \% $, when T$_{c}$ and T$_{h}$ are fixed at 500 and 1200 K, respectively. The value of $\eta$ indicates that Na$_{0.74}$Co$ _{0.95} $Nb$ _{0.05} $O$ _{2} $ compound can be used as a p-leg for making TEG at high temperature region.   

\end{document}